\documentclass[aps,a4paper,showpacs,twocolumn,pra,floatfix]{revtex4}

\usepackage{graphicx}
\usepackage[ansinew]{inputenc}
\usepackage{array}
\usepackage{color}
\usepackage{amsmath}
\usepackage{amsxtra}
\usepackage{amstext}
\usepackage{amssymb}
\usepackage{latexsym}
\usepackage{dsfont}
\usepackage{float}
\usepackage{eucal}
\begin{document}

\title{Considerations about the Aharonov-Anandan Phase for Time Independent Hamiltonians}

\author{P.-L. Giscard}
\affiliation{B2 Institute, Department of Physics and College of Optical
Sciences\\The University of Arizona, Tucson, Arizona 85721}

\date{\today}

\begin{abstract}
We present a method for calculating the Aharonov-Anandan phase for time-independent Hamiltonians that avoids the calculation of evolution operators. We compare the generic method used to calculate the Aharonov-Anandan phase with the method proposed here through four examples; a spin-$\frac{1}{2}$ particle in a constant magnetic field, an arbitrary infinite-sized Hamiltonian with two known eigenvalues, a Fabry-Perot cavity with one movable mirror and a three mirrors cavity with a slightly transmissive movable middle mirror. \end{abstract}


\maketitle
\section{Introduction}
\label{sec:intro}
The geometric phase, which was first recognized to arise from quantum systems undergoing a cyclic motion by M. V. Berry \cite{Berry1984}, as been studied in much detail, both from the mathematical point of view and for various physical systems. The topological nature of Berry's phase and the Aharonov-Anandan phase has been investigated since its discovery, yielding deep insights in the cyclic motion of quantum systems \cite{Simon1983}, \cite{Anandan1987}, \cite{Samuel1988}, \cite{Anandan1990}. Despite a good understanding of these geometric phases \cite{GeoPapers}, \cite{GeometricalPhases_Physics} and their use in various branches of physics, from quantum phase transitions \cite{ZhuReview2008} to quantum computation (see for example \cite{Ekert2000}, \cite{Zhu2005}), few studies have focused on algebraic methods to calculate them \cite{Mooremethod1990}, \cite{Monteolivamethod1994}. In fact the generic method still in use to calculate geometric phases was given in the pioneering article by Aharonov and Anandan \cite{AAPhase1987}.

The main goal of this paper is to propose an alternative method for calculating geometric phases arising from time independent Hamiltonians that avoids the explicit calculation of the evolution operator. We also derive results concerning the condition for an Hamiltonian to give rise to a cyclic motion of a state and on the possible resulting total phases (see Sec.\ref{subsec:calcgeo}). Sec.\ref{sec:exemgeo} then gives four examples of calculation of the geometric phase, showing that the results coincide with those obtained using the generic way to determine it. This also enables a comparison of the complexity involved in each method. Finally, in the last example we calculate the period and the total phase of a three-mirror cavity, using both the generic method and the new method proposed here. 

\section{Calculating the geometric phase}
\label{sec:geophase}

\subsection{A simplified expression for the geometric phase}
\label{subsec:calcgeo}

Let us first review the technique used to calculated the geometric phase arising from a cyclic motion of a quantum system.
A basic idea for the calculation of the geometric phase is to calculate the dynamical phase ($\varphi_{Dyn}$) and to subtract it from the total phase accumulated during one cyclic evolution of the system under consideration. In the following we only study the calculation of the Aharonov-Anandan phase (AA-phase).

Let $\tau$ be the period of this evolution and $\phi$ the phase difference between the initial state $|\Psi(0)\rangle$ and $|\Psi(\tau)\rangle$, i.e. $|\Psi(\tau)\rangle=e^{i\phi}|\Psi(0)\rangle$. 
Now let $f(t)$ be a continuous function of time so that $f(\tau)-f(0)=\phi$, and let $|\widetilde{\Psi(t)}\rangle=e^{if(t)}|\Psi(t)\rangle$. Aharonov and Anandan have proven in \cite{AAPhase1987} that the geometric phase $\gamma$ is given by 
\begin{equation}
\label{eq: geo}
\gamma=\int_{0}^{\tau}\langle\widetilde{\Psi(t)}|i\frac{d|\widetilde{\Psi(t)}\rangle}{dt}dt. 
\end{equation} 
For practical reasons, one uses instead the equivalent following formula that involves $|\Psi(t)\rangle$ explicitly
\begin{equation}
\label{eq: geo2}
\gamma=\phi+i\int_{0}^{\tau}\langle\Psi(t)|\dot{\Psi(t)}\rangle dt\equiv \phi-\varphi_{Dyn}, 
\end{equation} 
where one has to calculate $|\Psi(t)\rangle$ and then $|\dot{\Psi(t)}\rangle$ from the Hamiltonian, using the evolution operator $U(t)$. If for any time $t$ and $t'$, $[H(t),H(t')]=0$, then $U(t)=e^{-iHt/\hbar}$. However for complicated systems, calculating $U(t)$ is not trivial \cite{BOSE2MCStates1997}, \cite{BhattacharyaSqueeze2007} and the methods used for disentangling it are generally cumbersome \cite{WunscheDisentangle2002}, \cite{EchaveBCH1991}. Furthermore once a disentangled form of $U$ has been found, its action on a given initial state $|\Psi(0)\rangle$ may lead to involved expressions for $|\Psi(t)\rangle$. But as we show later on, these explicit calculations can be avoided if the Hamiltonian can be written as $H'(t)=H(t)+\textbf{1}a(t)$ where $a(t)$ is a function of time (not an operator) and $H(t)$ obeys $[H(t),H(t')]=0$.

As the geometric phase arising from a given Hamiltonian is gauge invariant (\cite{Berry1984}, \cite{AAPhase1987}), the geometric phases arising from $H'(t)$ and $H(t)$ are the same. 
Using $i|\dot{\Psi(t)}\rangle=\hbar^{-1}H(t)|\Psi(t)\rangle$ and $|\Psi(t)\rangle=U(t)|\Psi(0)\rangle$ the geometric phase is therefore 
\begin{equation}
\label{eq: geo3}
\gamma=\phi+\hbar^{-1}\int_{0}^{\tau}\langle\Psi(0)|U^{\dagger}(t)H(t)U(t)|\Psi(0)\rangle dt. 
\end{equation} 
For Hamiltonians that commute with themselves at any time, $[U(t'),H(t)]=[U^{\dagger}(t'),H(t)]=0$ and since $U^{\dagger}(t)U(t)=U(t)U^{\dagger}(t)=\textbf{1}$, the geometric phase reduces to 
\begin{equation}
\label{eq: geo4}
\gamma=\phi+\hbar^{-1}\int_{0}^{\tau}\langle\Psi(0)|H(t)|\Psi(0)\rangle dt. 
\end{equation} 
This expression can be further simplified in the case of time independent Hamiltonians to
\begin{equation}
\label{eq: geo5}
\gamma=\phi+\frac{\tau}{\hbar}\langle\Psi(0)|H|\Psi(0)\rangle.
\end{equation} 
Since the system evolves according to the Schr\"{o}dinger equation, that is to say at constant energy, and is driven by a time independent Hamiltonian, $\tau$ and $\phi$ are the same for any state located on the closed loop formed by the cyclic time-evolution in the projective Hilbert space. In other terms, starting the cyclic evolution from any point of the loop always yields the same geometric phase.

\subsection{Finding the Period and Total Phase of the Cyclic Motion}
\label{subsec:calcgeo}

\subsubsection{The Period $\tau$}
\label{subsubsec:period}
In the following we only consider time-independent Hamiltonians. The period $\tau$ and the phase $\phi$ are generally derived from physical considerations and a careful study of the evolution operator $U(t)$. 
One can however avoid such a study by working only with the Hamiltonian. Let $B=\{|\phi_{k}\rangle\}$ be a basis in which $H$ is diagonal and $\Lambda=\{\lambda_{k}\}$ the corresponding set of its eigenvalues. 
Let $B_{\Psi}\subseteq B$ be the smallest set of eigenvectors needed to decompose a state $|\Psi\rangle$ on $B$ and let $\Lambda_{\Psi}\subseteq\Lambda$ be the corresponding set of eigenvalues. Finally, let $\Delta E_{\Psi}$ be the set of \textit{non-zero} energy spacings in $\Lambda_{\Psi}$, i.e. $\Delta E_{\Psi}=\{\Delta E_{k,i}=\lambda_{k}-\lambda_{i}\}_{\lambda_{k,i}\in\Lambda_{\Psi},~\lambda_{k}\neq\lambda_{i}}$

Then, we show in the Appendix \ref{sec:apptau} that the period $\tau$ of the cyclic motion for $|\Psi\rangle$ is proportional to the least common multiple (LCM) of the inverse of $\Delta E_{\Psi}$. Rigorously we find 
\begin{equation}
\label{eq:tau}
\tau=2\pi\hbar ~\mathrm{LCM}\left(\Delta E_{\Psi}^{-1}\right).
\end{equation}

A simple example of application of Eq.(\ref{eq:tau}) is the calculation of the period of evolution of a coherent state driven by a free field Hamiltonian (e.g. $H=\hbar\omega a^{\dagger}a$). Eq.(\ref{eq:tau}) gives
\begin{equation}
\label{eq:taufreefield}
\tau=2\pi\hbar ~\mathrm{LCM}\left((\hbar\omega n)^{-1}\right)_{n\in \mathbb{Z^{*}}}=2\pi/\omega,
\end{equation} 
where we have used the fact that the LCM of all $1/n$ (with $n$ a non-zero integer) is 1.\\

\subsubsection{Conditions of Cyclicality}
The form of the period $\tau$ of Eq.(\ref{eq:tau}) provides a means to test if a given Hamiltonian will yield a cyclic evolution of a given initial state. Indeed, if the system initially in state $|\Psi\rangle$ effectively undergoes a cyclic motion, then $\tau$ must be finite, i.e. the least common multiple involved in Eq.(\ref{eq:tau}) must be finite.  

Therefore, there must be no two different elements of $\Delta E_{\Psi}^{-1}$ that are incommensurable. In the case where $\Lambda_{\Psi}$ contains more than two different eigenvalues \cite{2level}, this means that if one of the eigenvalues in $\Lambda_{\Psi}$ is irrational (say equal to $\kappa$), then the only way $H$ could yield a cyclic evolution of the state $|\Psi\rangle$ is that all the eigenvalues in $\Lambda_{\Psi}$ are rational multiples of $\kappa$.
 
Another condition for infinite Hamiltonians is that the set $\Delta E_{\Psi}^{-1}$ must be bounded in $\mathbb{R}$. This means that there must not be infinitely close eigenvalues in $\Lambda_{\Psi}$.

Remarkably, these conditions are always fulfilled if $\Lambda_{\Psi}$ contains exactly two different eigenvalues. 

\subsubsection{Total Phase $\phi$}
\label{subsubsec:totalphase}
In the same way than for $\tau$, we show in Appendix \ref{sec:apptau} that the total phase accumulated by an initial state $|\Psi\rangle$ after one cycle of evolution is given by 
\begin{equation}
\label{eq:phi}
\phi=2\pi\left[n-\lambda~\mathrm{LCM}\left(\Delta E_{\Psi}^{-1}\right)\right],
\end{equation}
where $n$ is an integer that depends on $\lambda\in\Lambda_{\Psi}$. This expression being valid for any eigenvalue of $\lambda$ of $\Lambda_{\Psi}$, as soon as $0\in\Lambda_{\Psi}$, then the total phase must be $\phi=2\pi$.

In the case where $\Lambda_{\Psi}$ contains more than two different eigenvalues, Eq.(\ref{eq:phi}) also yields a more general condition on $\phi$. Indeed, as we have seen before, the least common multiple involved in the above equation must be finite for the system to exhibit a cyclic motion of the state $|\Psi\rangle$. 
Thus in Eq.(\ref{eq:phi}), $\lambda~\mathrm{LCM}\left(\Delta E_{\Psi}^{-1}\right)$ is always rational as soon as the system exhibits a cyclic motion, so that
\begin{equation}
\label{eq:phi3}
\frac{\phi}{\pi}~\mathrm{is~rational}.
\end{equation}
This condition does not holds if $\Lambda_{\Psi}$ contains exactly two different eigenvalues (in which case $\phi/\pi$ can be irrational).

\subsubsection{A Method for Time Independent Hamiltonians}
\label{subsubsec:meth}
Ideally the geometric phase is just given by Eq.(\ref{eq: geo5}). Nevertheless, as we see from Eq.(\ref{eq:tau}) and Eq.(\ref{eq:phi}), $\tau$ and $\phi$ require a complete knowledge of the spectrum $\Lambda_{\Psi}$ involved in the decomposition of $|\Psi\rangle$ on the base $B$. 
It is however generally possible to use or impose relations on $H$ so that the least common factor is easier to calculate (such assumptions are often made in the generic method, see e.g. Ref.\cite{ZHUGeo2MC1998}).

In the case where not all the eigenvalues are known, and there is no relation that one can or wants to impose on $H$ to calculate the AA-phase, it is still possible to partially know it from only one non-zero element of $\Lambda_{\Psi}$. If the only known eigenvalue is zero, we have no information on $\tau$ but $\phi=2\pi$ is completely determined. 

Let $\lambda\in\Lambda_{\Psi}$. As shown in Appendix \ref{sec:apptau}, there exists an integer $n$ such that 
\begin{equation}
\label{eq:condcond}
\hbar^{-1}\lambda \tau=-\phi+2n\pi.
\end{equation} Performing the gauge transformation $H'=H+\frac{2n\pi\hbar}{\tau}$ which leaves the AA-phase unchanged, Eq.(\ref{eq:condcond}) becomes $\hbar^{-1}\lambda\tau=-\phi$ and Eq.(\ref{eq: geo5}) transforms to
\begin{eqnarray}
\label{eq: resultat}
\gamma&=&\phi\left[1-\lambda^{-1}\langle\Psi(0)|H|\Psi(0)\rangle\right]+2\pi n,
\end{eqnarray}
which is consistent with the gauge invariance of the geometric phase. Note that the total phase $\phi$ in that equation is unknown.
It is also possible to recast the expression of the AA-phase in term of an unknown period $\tau$ as
\begin{eqnarray}
\label{eq: resultatau}
\gamma&=&\frac{\tau}{\hbar}\left[\langle\Psi(0)|H|\Psi(0)\rangle-\lambda\right]+2\pi n.
\end{eqnarray}

These relations hold for any element of $\Lambda_{\Psi}$. 
Only the total phase or the period remains unknown. However, if another element of $\Lambda_{\Psi}$ is known, it is possible to constrain the possible total phases or periods using Eq.(\ref{eq:phi}) for the two eigenvalues, see Sec.\ref{subsec: infinitesize}. In the opposite way, if the total phase or the period is known from physical considerations or if $0\in\Lambda_{\Psi}$, the geometric phase is given by the knowledge of only one non-zero eigenvalue and it is possible to constrain the unknown part of the spectrum of the considered Hamiltonian. These procedures are detailed in Sec.\ref{sec:exemgeo}.

\section{Some examples}
\label{sec:exemgeo}

In this section, we illustrate the calculation of the Aharonov-Anandan phase for various physical systems. We assume that the considered systems effectively undergo cyclic motions when starting in the proposed states $|\Psi(0)\rangle$.

\subsection{Spin-$\frac{1}{2}$}
\label{subsec: spin}
Consider first the precession of a spin-$\frac{1}{2}$ around a constant magnetic field $B_{0}$. 
The Hamiltonian of that system in the rest frame is 
\begin{equation}
\label{eq: hamilspin}
H=-\mu B_{0} \sigma_{z}
\end{equation}
where $\mu$ is the magnetic moment of the particle and $\sigma_{z}$ is the Pauli matrix
\begin{equation}
\label{eq:pauli} \sigma_{z} =
\begin{pmatrix}
1 &  0\\
0 & -1  
\end{pmatrix}.
\end{equation}
Now let 
\begin{equation}
\label{eq:psinot} |\Psi(0)\rangle =
\begin{pmatrix}
\mathrm{cos}(\theta/2)\\
\mathrm{sin}(\theta/2)  
\end{pmatrix}.
\end{equation}
The generic derivation of the AA-phase for this system is given in numerous papers and books (see \cite{AAPhase1987}) and yields the famous result $\gamma=\pm\pi(1-\mathrm{cos}(\theta))$, the total phase $\phi=\pm \pi$ being found from physical considerations.\\

Alternatively since $\Lambda_{\Psi}$ (and in fact $\Lambda$) is known, one can calculate $\tau$ from Eq.(\ref{eq:tau}), $\tau=2\pi\hbar \mathrm{LCM}(\frac{1}{\mu B_{0}-(-\mu B_{0})})=\frac{\pi\hbar}{\mu B_{0}}$, 
so that $\tau \hbar^{-1} \langle\Psi(0)|H|\Psi(0)\rangle = -\pi \mathrm{cos}(\theta)$.
Moreover, Eq.(\ref{eq:phi}) entails that the only possible total phase is solution of $\phi=2\pi (n-\frac{1}{2})=2\pi(m+\frac{1}{2})$ with $n$ and $m$ two different integers, that is $\phi=\pm\pi$. We thus recover $\gamma=\pi(1-\mathrm{cos}(\theta))$.\\

It is also possible to calculate the AA-phase from only one element of $\Lambda_{\Psi}$ using the fact that $\phi=\pm\pi$ is known from physical considerations. We know that $\langle \Psi(0)|H|\Psi(0)\rangle = -\mu B_{0} \mathrm{cos}(\theta)$. If we choose the eigenvalue $\lambda_{+}=\mu B_{0}$, Eq.(\ref{eq: resultat}) gives us that $\gamma=\phi(1+cos(\theta))=\pm\pi(1+cos(\theta))=\pm\pi\mp\mathrm\pi{cos}(\theta)=\mp\pi\mp\mathrm\pi{cos}(\theta)+2\pi\equiv\pm\pi(1-\mathrm{cos}(\theta))$ as it should. If instead we use, $\lambda_{-}=-\mu B_{0}$, we have likewise immediately that $\gamma=\pm\pi(1-\mathrm{cos}(\theta))$.

\subsection{Infinite-size arbitrary Hamiltonian}
\label{subsec: infinitesize}

Consider now an infinite Hamiltonian and a state $|\Psi\rangle$ so that we know two different non-zero elements $\Lambda_{1}$ and $\Lambda_{2}$ of $\Lambda_{\Psi}$. We further assume that this Hamiltonian yields a cyclic motion of the state $|\Psi\rangle$. Our goal is to show how one can derive the geometric phase and constrain the possible total phases, periods and the unknown part of the spectrum using these two eigenvalues. Further informations such as the knowledge of the period from physical considerations or experimental results allow to refine these constraints.
The Hamiltonian is given by the infinite matrix \cite{position}
\begin{equation}
\label{eq:hamilH} H =
\begin{pmatrix}
\Lambda_{1} &  0 & 0& 0&\ldots \\
0&\Lambda_{2}&0&0&\ldots\\
0 & 0&B_{1} & B_{2}&\ldots \\
0& 0&C_{1}& C_{2}&\ldots\\
\vdots&\vdots&\vdots&\vdots&\ddots\\
\end{pmatrix}
\end{equation}
The geometric phase is given by Eq.(\ref{eq: geo5})
\begin{eqnarray}
\label{eq:ex}
\gamma=\phi+\frac{\tau}{\hbar}\langle H\rangle
\end{eqnarray}
where $\langle H\rangle$ indicates the expectation value of $H$ calculated for any state $|\Psi(t)\rangle$ located on the closed loop formed by the cyclic time-evolution in the projective Hilbert space.\\ 

Consider first the most general case where none of the total phase and the period are known. Using Eq.(\ref{eq:phi}), we can constrain $\phi$ through the relations 
\begin{equation}
\label{eq:phiconstraint}
\phi=2\pi n-\Lambda_{1}\frac{\tau}{\hbar}=2\pi m-\Lambda_{2}\frac{\tau}{\hbar}.
\end{equation}
Using the first relation, we find $\tau$ as a function of $\phi$ which allows to find $\phi$ by introducing the expression of $\tau$ in the second relation. We find 
\begin{eqnarray}
\phi&=&2\pi\frac{\Lambda_{1}m-\Lambda_{2}n}{\Lambda_{1}-\Lambda{2}},\label{eqn:condphii}\\
\tau&=&2\pi\hbar\frac{n-m}{\Lambda_{1}-\Lambda_{2}}.\label{eqn:condtaui}
\end{eqnarray}
Now let us change of gauge to get $\phi=0$ and thus $\Lambda_{1}m=\Lambda_{2}n$. This does not change the period nor the geometric phase and only shifts the whole spectrum of the Hamiltonian without affecting the dynamics of the system. The period is now expressed as
\begin{equation}
\label{eq:condtau}
\tau=2\pi\hbar\frac{n}{\Lambda_{1}}.
\end{equation}
This gives the geometric phase 
\begin{eqnarray}
\gamma&=&2\pi \frac{n}{\Lambda_{1}}\langle H\rangle[2\pi],\label{eqn:gammal1}\\
\gamma&=&2\pi \frac{m}{\Lambda_{2}}\langle H\rangle[2\pi].\label{eqn:gammal2}
\end{eqnarray}
As the geometric phase is defined modulo $2\pi$, if $\langle H\rangle/\Lambda_{1,2}$ is rational \cite{ratiostate}, there is only a finite number of values of $n$ and $m$ that will yield different results up to $2\pi$ phase factors.
It is interesting to remark that Eqs.(\ref{eq:phi3}) and (\ref{eqn:condphii}) imply that $\Lambda_{1}/\Lambda_{2}$ is a rational number, which, as we know from Sec.\ref{subsubsec:period} and Sec.\ref{subsubsec:totalphase}, is consistent with the assumed fact that the system undergoes a cyclic motion. 

We now turn to the unknown part of the spectrum. Any eigenvalue of $\Lambda_{\Psi}$ must fulfill and equation similar to Eq.(\ref{eq:phiconstraint}) and can thus be constrained as $\phi$ is now known to be $0$. This leads to
\begin{equation}
\Lambda_{k}n=\Lambda_{1}k'
\end{equation}
$k'$ being the integer entering Eq.(\ref{eq:phiconstraint}) when written for $\Lambda_{k}$. 
Such a procedure can be carried using other eigenvalues if some are known beyond $\Lambda_{1}$ and $\Lambda_{2}$. This will lead to new relations on the unknown $\Lambda_{k}$, $\phi$ and $\tau$ which further constrain their possible values.

This illustrates that it is possible to partially know the geometric phase, $\phi$, $\tau$ and the unknown part of $\Lambda_{\Psi}$ from an incomplete spectrum from the condition of periodicity of the considered state $|\psi\rangle$. If a total knowledge of the geometric phase is required, it is also possible to directly impose constrains on the spectrum of $H$, as we now illustrate.

\subsection{Fabry-Perot Cavity with a Movable Mirror}
\label{subsec: 2MC}
In this section we consider the problem of a single mode of the light field inside a Fabry-Perot cavity with one movable mirror. This system has been studied both theoretically and experimentally in great detail in the context of the emerging field of cavity optomechanics. It is characterized by a large variety of quantum mechanical features, such as non classical states \cite{BOSE2MCStates1997} or entanglement \cite{VitaliVib2007}, \cite{BhattacharyaRot2007}, \cite{LauratCMexp2005}. 

This system is described to an excellent approximation by the Hamiltonian \cite{ManciniHamil1997}
\begin{equation}
\label{eq: hamil2MC}
H=\hbar \omega_{f}a^{\dagger}a+\hbar \omega_{m}b^{\dagger}b-\hbar g a^{\dagger}a(b+b^{\dagger}),
\end{equation}
where $a^{\dagger}$ ($a$) is the bosonic creation (annihilation) operator for the cavity field mode. Similarly, the mirror is treated as a quantum harmonic oscillator of frequency $\omega_{m}$ with $b^{\dagger}$ ($b$) the bosonic creation (annihilation) operator, and $g$ is the opto-mechanical coupling.\\

We first compute the geometric phase in this system via the generic method. The evolution operator $U(t)$ was calculated and disentangled in Ref.\cite{BOSE2MCStates1997} as
\begin{eqnarray}
\label{eq:Udisentangle}
\begin{array}{lll}
U(t)&=&\mathrm{exp}[-ira^{\dagger}a\omega_{m}t]\mathrm{exp}[ik^{2}(a^{\dagger}a)^{2}(\omega_{m}t-\mathrm{sin}~\omega_{m}t)]\nonumber\\
&&\times\mathrm{exp}[ka^{\dagger}a(\eta b^{\dagger}-\eta^{*}b)]\mathrm{exp}[-ib^{\dagger}b\omega_{m}t],
\end{array}
\end{eqnarray}
where $r=\omega_{f}/\omega_{m}$, $k=g/\omega_{m}$ and $\eta=1-e^{-i\omega_{m}t}$. Consider the initial state 
\begin{equation}
\label{eq:psizero}
|\Psi_{0}\rangle=\sum_{n=0}^{\infty}C_{n}|n\rangle_{f}\otimes|\beta\rangle_{m},
\end{equation}
where $|n\rangle_{f}$ is a Fock state of the cavity field and $|\beta\rangle_{m}$ a coherent state of the mirror. Following \cite{ZHUGeo2MC1998}, the state $|\Psi(t)\rangle$ at time $t$ is 
\begin{widetext}
\begin{equation}
\label{eq:psit}
|\Psi(t)\rangle=\sum_{n=0}^{\infty}C_{n}\mathrm{exp}[-irn\omega_{m}t+ik^{2}n^{2}(\omega_{m}t-\mathrm{sin}~\omega_{m}t)]\mathrm{exp}[\frac{1}{2}kn(\eta\beta^{*}e^{i\omega_{m}t}-\eta^{*}\beta e^{-i\omega_{m}t})]|n\rangle_{f}|\beta e^{-i\omega_{m}t}+kn(1-e^{-i\omega_{m}t})\rangle_{m}\nonumber
\end{equation}
\end{widetext}
As in Ref.\cite{ZHUGeo2MC1998}, we assume that $r$ is an integer and we choose $k$ so that $k^{2}\tau=2q\pi$ and $\tau=2p\pi$, where $p$ and $q$ are the smallest integers satisfying $k^{2}=q/p$ (this is possible by constraining for example the length $L$ of the cavity). In that case the system's motion is periodic with total phase $\phi=2\pi$. 

We still have to calculate $d|\Psi(t)\rangle/dt$ from Eq.(\ref{eq:psit}) and compute Eq.(\ref{eq: geo2}). 
The resulting geometric phase, calculated in Ref.\cite{ZHUGeo2MC1998}, is
\begin{equation}
\label{eq: geo2mc}
\gamma=2\pi\left[1+\frac{p}{\omega_{m}}(r-2kRe(\beta))\langle\tilde{n}\rangle_{f}+\frac{p}{\omega_{m}}|\beta|^{2}\right],
\end{equation}
with $\langle\tilde{n}\rangle_{f}=\sum_{n=0}^{\infty}n|C_{n}|^{2}$.\\ 

We now rederive that same result using our new method. We remark that $0\in\Lambda_{\Psi}$ is an eigenvalue of $H$ with eigenvector $|0\rangle_{\mathrm{field}}|0\rangle_{\mathrm{mirror}}=|00\rangle$, thus $\phi=2\pi$. Furthermore, using the initial state of Eq.(\ref{eq:psizero}), the Hamiltonian of Eq.(\ref{eq: hamil2MC}) allows us to compute easily
\begin{equation}
\label{eq:HamilMinValue}
\langle\Psi(0)|H|\Psi(0)\rangle=\hbar\omega_{m}\left[(r-2kRe(\beta))\langle\tilde{n}\rangle_{f}+|\beta|^{2}\right],\nonumber
\end{equation}
and therefore the most general expression for the geometric phase is 
\begin{equation}
\label{eq:geogen}
\gamma=2\pi+\omega_{m}\tau(r-2kRe(\beta))\langle\tilde{n}\rangle_{f}+\omega_{m}\tau|\beta|^{2}.
\end{equation}
The remaining difficulty is to find the period $\tau$. But we can remark also that any state of the form $|0n\rangle$, where $n$ is a positive integer, is eigenstate of $H$ with eigenvalue $\hbar\omega_{m}n\in\Lambda_{\Psi}$. Thus Eq.(\ref{eq:phi}) together with $\phi=2\pi$ gives us directly that $\hbar\omega_{m}\mathrm{LCM}\left(\Delta E_{\Psi}^{-1}\right)=p$ is a non-zero integer. Thus $\tau=\frac{2\pi\hbar}{\hbar\omega_{m}}p=2\pi\frac{p}{\omega_{m}}$. Therefore we obtain from Eq.(\ref{eq:geogen})
\begin{equation}
\label{eq:geogenresult}
\gamma=2\pi\left[1+\frac{p}{\omega_{m}}(r-2kRe(\beta))\langle\tilde{n}\rangle_{f}+\frac{p}{\omega_{m}}|\beta|^{2}\right].
\end{equation}
This expression holds without any assumption on the parameters. However we want to check if we can make approximations similar to those performed in the generic method, so that we get Eq.(\ref{eq: geo2mc}) and Eq.(\ref{eq:geogenresult}) equal.

First we calculate $k^{2}\tau=2\pi\frac{r^{2}\hbar}{2mL^{2}\omega_{m}^{2}}p$. Now let us suppose that $r$ is an integer. As in the generic method, we constrain $L$ such that $\frac{\hbar}{2mL^{2}\omega_{m}^{2}}$ is a non-zero integer. This way, we see that there exists a non-zero integer $q$ such that $k^{2}=q/p$ and by definition of $\tau$, these integers are the smallest satisfying this equation. Thus with the assumptions that $r$ is an integer and that $L$ is constraint so that $\frac{\hbar}{2mL^{2}\omega_{m}^{2}}$ is a non-zero integer, we obtain that Eq.(\ref{eq: geo2mc}) and Eq.(\ref{eq:geogenresult}) are equal.

\subsection{Three-Mirror Fabry-Perot Cavity with One Movable Mirror}
\label{subsec: 3MC}
The last example that we consider is a three-mirror cavity with a slightly transmissive and movable middle mirror of mass $m$ and natural vibration frequency $\omega_{m}$. This system has been discussed in Refs.\cite{ThompsonExp3MC2008}, \cite{Bhattacharya3MC2007} and \cite{BhattacharyaCoolTrap2007}.

\subsubsection{Generic Method}
\label{subsubsec:evogeneric3mc}
Ref. \cite{Bhattacharya3MC2007} identifies a regime of coupling between the middle mirror and the left and right cavity fields that is linear in the position of the mirror for one field and quadratic for the other.
Expressing the momentum $p$ and the position $q$ of the mirror in terms of bosonic creation and annihilation operators $c^{\dagger}$ and $c$ as $p= i\sqrt{2m\hbar\omega_{m}}(c^{\dagger}-c)$ and $
q=\sqrt{\frac{\hbar}{2m\omega_{m}}}(c^{\dagger}+c)$ the Hamiltonian is 
\begin{widetext}
\begin{eqnarray}
\label{eq:Hq}
H =\hbar \omega_D a^{\dagger} a +
\hbar \omega_S b^{\dagger} b +\hbar C_{D}a^{\dagger}a(c+c^{\dagger})+(\hbar \omega_{m}+\hbar C_{S}b^{\dagger}b) c^{\dagger}c+\frac{\hbar C_{S}}{2}b^{\dagger}b(1+c^{2}+c^{\dagger,2}),
\end{eqnarray}
\end{widetext}
where $a^{\dagger}$ ($a$) and $b^{\dagger}$ ($b$) are the bosonic creation (annihilation) operators of the light fields in, respectively, the left and the right cavity, and
\begin{eqnarray}
\label{eq:Hconstants}
C_{D}&=&\frac{\xi_D}{\sqrt{2m\omega_m/\hbar}}, \nonumber \\
C_{S}&=&\frac{\hbar \xi_S}{m \omega_m}, \nonumber \\
\end{eqnarray}
with $\xi_{D}$ and $\xi_{S}$ constants, whose explicit forms are given in Ref.\cite{BhattacharyaSqueeze2007}. We further introduce
\begin{eqnarray}
\label{eq:Hconstants2}
\chi &=&\left[\omega_m(\omega_m + 2 C_Sb^{\dagger}b)\right]^{1/2},\nonumber \\
\delta&=&(\omega_{D}-C_{D})a^{\dagger}a+\omega_{s}b^{\dagger}b,\nonumber \\
 \nu &=&
    C_{D}a^{\dagger}a\chi^{-1}\left[\omega_m\chi^{-1}\left (\cos\chi
    t-1\right)-i\sin \chi t \right], \nonumber \\
    |\kappa|&=&\left|\sinh^{-1}
    \left(C_{S}b^{\dagger}b\chi^{-1} \sin \chi t \right)\right|.
\end{eqnarray}
Note that using realistic parameters for the system (see \cite{BhattacharyaSqueeze2007}), one obtain $\langle\chi\rangle\simeq\omega_{m}$. 

Using a semi-classical approximation for the light fields, the evolution operator corresponding to Eq.(\ref{eq:Hq}) was disentangled in Ref.\cite{BhattacharyaSqueeze2007}. A full quantum treatment leads to a similar evolution-operator
\begin{equation}
\label{eq:evoquadratic}
U(t)=e^{i\delta t}D(\nu)R(\Phi)S(\kappa)
\end{equation}
where $D(\nu)=\mathrm{exp}[\nu c^{\dagger}-\nu^{*} c]$ is the displacement operator, $R(\Phi)=\mathrm{exp}(i\frac{\Phi}{4}(c^{\dagger}c+cc^{\dagger}))$ is a rotation operator, and $S(\kappa)=\mathrm{exp}[\kappa^{*}c^{2}/2-\kappa c^{2,\dagger}/2]$ is a squeezing operator. It has been shown in Ref.\cite{BhattacharyaSqueeze2007} that the rotation due to $R$ exactly oppose that due to $S$, so that Eq.(\ref{eq:evoquadratic}) is equivalent to
\begin{equation}
\label{eq:evoquadraticsimple}
U(t)=e^{i\delta t}D(\nu)S(|\kappa|)
\end{equation}
Now using the expressions of $\delta$, $\nu$ and $\kappa$ from Eq.(\ref{eq:Hconstants2}), one sees that $U(t)$ is periodic with period $\tau\simeq\frac{2\pi}{\omega_{m}}$. For that value of $\tau$, $U(\tau)=U(0)$, so that $\phi=2\pi$.
This shows the difficulty of working with the generic method. Indeed, obtaining the geometric phase directly from $U(t)$ leads to rather involved expressions for both $|\Psi(t)\rangle$ and $d|\Psi(t)\rangle/dt$.
 
\subsubsection{Obtaining the geometric phase}
\label{subsubsec:evogeneric3mceff}
The results for $\tau$ and $\phi$ obtained in the previous section can also be obtained from $H$ using a  simple evaluation of some of its eigenvalues, i.e. without calculating $U(t)$. Consider for example an initial product state 
\begin{equation}
\label{eq:psi03mc}
|\Psi(0)\rangle=\sum_{n=0}^{\infty}A_{n}|n\rangle_{a}\sum_{n=0}^{\infty}B_{n}|n\rangle_{b}\sum_{n=0}^{\infty}M_{n}|n\rangle_{m}.
\end{equation}
First, we remark that the state  $|0\rangle_{\mathrm{field~a}}|0\rangle_{\mathrm{field~b}}|0\rangle_{\mathrm{mirror}}=|000\rangle$ is an eigenstate of $H$ with eigenvalue $0\in\Lambda_{\Psi}$. Thus we have $\phi=2\pi$. Furthermore the state $|00n\rangle$ (where $n$ is a positive integer) is eigenstate of $H$ with eigenvalue $\hbar\omega_{m}n\in\Lambda_{\Psi}$. Therefore there exists a non-zero integer $p$ such that $\tau=2\pi\frac{p}{\omega_{m}}$. This gives the period of cyclic motion without any assumption. It is nevertheless possible to find specific values of $p$ using approximations. For example, considering that $\langle2C_{S}b^{\dagger}b\rangle\ll\omega_{m}$ and $\langle C_{D}a^{\dagger}a\rangle\ll \omega_{m}$, $\omega_{D}/\omega_{m}$ and $\omega_{S}/\omega_{m}$ integers, yields $p=1$ (free field of frequency $\omega_{m}$).

To calculate the geometric phase, it remains to introduce $\tau$ and $\phi$ in Eq.(\ref{eq: geo5}) and to choose an initial state. The expectation value of $H$ at $t=0$ is easily calculated from Eq.(\ref{eq:Hq}) and Eq.(\ref{eq:psi03mc}), and the most general expression for the geometric phase given by Eq.(\ref{eq: geo5}) is
\begin{widetext}
\begin{eqnarray}
\label{eq: geoquadratic}
\gamma&=&2\pi\left[1+p\left(\frac{\omega_{D}}{\omega_{m}}\langle\tilde{a}\rangle+\frac{\omega_{S}}{\omega_{m}}\langle\tilde{b}\rangle+2\frac{C_{D}}{\omega_{m}}\langle\tilde{a}\rangle\sum_{n=0}^{\infty}\sqrt{n+1}Re(M_{n}M_{n+1})+\ldots\right.\right.\nonumber\\
&&\ldots+\left.\left.\langle\tilde{m}\rangle+\frac{C_{S}}{\omega_{m}}\langle\tilde{b}\rangle\langle\tilde{m}\rangle+\frac{C_{S}}{2\omega_{m}}\langle\tilde{b}\rangle\left(1+2\sum_{n=0}^{\infty}\sqrt{(n+2)(n+1)}Re(M_{n}M_{n+2})\right)\right)\right].
\end{eqnarray}
\end{widetext}
Eq.(\ref{eq: geoquadratic}) is further simplified if we consider the initial state to be a product of coherent states $|\Psi(0)\rangle=|\alpha\rangle_{a}|\beta\rangle_{b}|\mu\rangle_{m}$, in which case
\begin{widetext}
\begin{equation}
\label{eq:geosimple}
\gamma=2\pi\left[1+p|\alpha|^{2}\left(\frac{\omega_{D}}{\omega_{m}}+2\frac{C_{D}}{\omega_{m}}Re(\mu)\right)+p|\beta|^{2}\left(\frac{\omega_{S}}{\omega_{m}}+\frac{C_{S}}{\omega_{m}}(\frac{1}{2}+2Re(\mu)^2)\right)+p|\mu|^{2}\right].
\end{equation}
\end{widetext}

\section{Conclusion}
\label{sec:conclu}

We have shown how, for time independent Hamiltonians, it is possible to derive the geometric phase arising from a cyclic motion of the system without calculating the evolution operator. We have proposed several ways to compute $\tau$, $\phi$ and the resulting geometric phase $\gamma$ respectively through Eq.(\ref{eq:tau}), Eq.(\ref{eq:phi}), Eq.(\ref{eq: geo5}) and Eq.(\ref{eq: resultat}). This led to the conditions for an Hamiltonian to yield a cyclic motion of a given state.
We have further shown how using the periodicity condition and knowing only a part of the spectrum of the Hamiltonian it is possible to derive the geometric phase, with the remaining uncertainty contained in the total phase or the period of the cyclic motion. This also permits to build constraints on the unknown part of the spectrum of $H$.
Through the $\mathrm{spin}-\frac{1}{2}$ particle and the two mirror vibrating Fabry-Perot cavity examples, we have shown that our method gives the same result as the generic method.

\begin{acknowledgments}
The author thanks M. Bhattacharya for starting his interest in geometric phases and P. Meystre for many helpful discussions. 
This work is supported in part by the US Office of Naval Research,
by the National Science Foundation, and by the US Army Research
Office. 
\end{acknowledgments}

\appendix
\section{Expressions of $\tau$ and $\phi$}
\label{sec:apptau}
In this Appendix we derive the expressions for $\tau$ and $\phi$ respectively given in Eq.(\ref{eq:tau}) and Eq.(\ref{eq:phi}).

Let $B=\{|\phi_{k}\rangle\}$ be a basis in which $H$ is diagonal and $\Lambda=\{\lambda_{k}\}$ the corresponding set of its eigenvalues. 
Let $B_{\Psi}\subseteq B$ the be smallest set of eigenvectors needed to decompose a state $|\Psi\rangle$ on $B$ and let $\Lambda_{\Psi}\subseteq\Lambda$ be the corresponding set of eigenvalues. Finally, let $\Delta E_{\Psi}$ be the set of \textit{non-zero} energy spacings in $\Lambda_{\Psi}$, i.e. $\Delta E_{\Psi}=\{\Delta E_{k,i}=\lambda_{k}-\lambda_{i}\}_{\lambda_{k,i}\in\Lambda_{\Psi}}$

In the basis $B$, $U(t)$ is diagonal and the elements of the diagonal have the form $\mathrm{exp}[-i\hbar^{-1}\lambda t]$. By definition of the period, $\tau$ is given by 
\begin{equation}
\label{eq:deftau}
\tau=\mathrm{min}_{t_{0}}\{t_{0}\neq0,~U(t_{0})|\Psi\rangle=e^{i\phi}U(0)|\Psi\rangle\}, 
\end{equation}
which, in the diagonal basis, means that there exists $t_{0}$ such as for all the elements $\lambda\in\Lambda_{\Psi}$, there exists an integer $n$ such that 
\begin{equation}
\label{eq: condition}
\hbar^{-1}\lambda t_{0}=-\phi+2n\pi
\end{equation}
and $\tau=\mathrm{min}_{t_{0}}\{t_{0}\}$. Let us say that the above condition has at least one non-zero solution, i.e. that the system effectively undergoes a cyclic evolution with total phase $\phi$.
We now subtract Eq.(\ref{eq: condition}) for two different elements of $\Lambda_{\Psi}$ (we suppose that $\Lambda_{\Psi}$ has at least two different eigenvalues, otherwise the state does not evolve with time). This leads to the following condition that is true for any pairwise different eigenvalues $\lambda_{k}$ and $\lambda_{i}$ elements of $\Lambda_{\Psi}$
\begin{equation}
\exists m\in\mathbb{Z}~\mathrm{such~that} ~\hbar^{-1}t_{0}\Delta E_{k,i}=2m\pi,
\end{equation} 
one can see that any $t_{0}/(2\pi\hbar)$ solution of Eq.(\ref{eq: condition}) must be proportional to a multiple of any $\Delta E_{k,i}^{-1}$ and therefore be among the common multiples of all elements of $\Delta E_{\Psi}^{-1}$. $\tau/(2\pi\hbar)$ being the smallest quantity to fulfill this condition, $\tau$ is proportional to the least common multiple (LCM) of $\Delta E_{\Psi}^{-1}$. We obtain Eq.(\ref{eq:tau}) given in Sec.\ref{subsubsec:period}
\begin{equation}
\label{eq:tauappendix}
\tau=2\pi\hbar ~\mathrm{LCM}\left(\Delta E_{\Psi}^{-1}\right)
\end{equation}
Note that in the case where $\phi$ is a integer multiple of $2\pi$, Eq.(\ref{eq:tauappendix}) reduces $\tau=2\pi\hbar~\mathrm{LCM}(\lambda^{-1})$, $\lambda\in\Lambda_{\Psi}$, as Eq.(\ref{eq: condition}) shows that for all $\lambda\in\Lambda_{\Psi}$, there exists an integer $m$ such that $t_{0}=2m\pi\hbar\lambda^{-1}$. Similarly, in the case where all the elements of $\Lambda_{\Psi}$ are equal (say to $\lambda$), $\tau$ simply reduces to $2\pi\hbar\lambda^{-1}$ (period of the complex exponential $\mathrm{exp}[-i\hbar^{-1}\lambda t]$.\\

As $\tau$ is known, we can derive the expression of the total phase $\phi$ inserting Eq.(\ref{eq:tauappendix}) in Eq.(\ref{eq: condition}):
\begin{equation}
\label{eq:phi2}
\phi=2\pi\left[n-\lambda~\mathrm{LCM}\left(\Delta E_{\Psi}^{-1}\right)\right]
\end{equation}
where $n$ is an integer and depend on the considered eigenvalue $\lambda\in\Lambda_{\Psi}$. Note that this expression is valid for any eigenvalue element of $\Lambda_{\Psi}$. 

\bibliography{AA-phase}

\begin{thebibliography}{28}
\bibitem[{2le()}]{2level}
\bibinfo{note}{As the sign of $\Delta E_{k,i}$ plays no role in the
  determination of the least common multiple, when $\Lambda_{\Psi}$ contains
  only two different eigenvalues there is only one possible $\Delta
  E^{-1}=\Delta E_{\Psi}^{-1}$, this quantity being of course never
  incommensurable to itself.}
\bibitem[{pos()}]{position}
\bibinfo{note}{The position of $\Lambda_{1}$ and $\Lambda_{2}$ in the matrix is
  arbitrary and has no incidence on the presented method.}
\bibitem[{rat()}]{ratiostate}
\bibinfo{note}{This condition is met as soon as all the $|\langle
  \phi_{k}|\psi\rangle|^{2}$ are rational numbers, $|\phi_{k}\rangle\in
  B_{\Psi}$ being an the eigenvector of $H$.}
\end{thebibliography}
\end{document}